Quantum propulsion and trapping of nano-objects by inhomogeneity-induced lateral Casimir forces

F. Bao, K. Shi, and S. He*

Lateral Casimir force near a laterally-inhomogeneous plate is first revealed by both rigorous simulations and proximity approximations. The inhomogeneity-induced lateral Casimir force provides a novel method to control the lateral motion of nano-objects above the plate, and makes source-free manipulations of them possible. When incorporated with the Casimir repulsion in a fluid, the lateral Casimir force is shown to dominate over Brownian motion and enables long-distance quantum propulsion and firm quantum trapping of nano-objects. Gratings of varying filling factors to mimic micro-scale inhomogeneity also confirm those effects. The idea to design asymmetric distributions of nano-structures paves the way to sophisticated tailoring of the lateral Casimir force.

The Casimir force, resulting from the variation of the free energy (including the zero-point energy contributions) when fluctuating electromagnetic (EM) fields are perturbed by material boundaries, is becoming increasingly important in micro- or nano-scale systems [1]. As an inter-molecular force between condensed objects, the Casimir force is an extension of the van der Waals force, taking into account both the retardation effect and the multiple scattering effect [2], with the consequence that it gets much stronger and is of relatively long range, usually extending to a distance of sub-microns. The above features on one hand make the Casimir force annoying in micro-electromechanical systems since it causes vacuum-separated, electrically neutral, and non-magnetic nano-objects to strongly interact with each other. On the other hand this also makes it a promising technique in non-touching applications, such as quantum actuation [3], quantum levitation [4], quantum ratchet [5], and, as we shall show in this letter, quantum propulsion and quantum trapping of nano-objects.

Relying on the reflectivity of fluctuating EM fields at boundaries, the Casimir force shows significant material and shape dependences. Homogeneous media are most intensively involved in relevant studies. In contrast to the common form (monotonic and attractive) of the Casimir force between two parallel perfect mirrors, an arrangement of materials would immediately flip the sign of the force from attraction to repulsion when the permittivities for two objects, 1 and 2, immersed in a fluid, 3, satisfy the so-called Dzyaloshinskii-Lifshitz-Pitaevskii (DLP) condition $(\varepsilon_1 - \varepsilon_3)(\varepsilon_2 - \varepsilon_3) < 0$ for a wide frequency range [4]. On the other hand, macroscopic objects made of anisotropic materials or fabricated with corrugations have preferable angular and/or lateral alignments, according to the lowest energy principle, and thus torques [6] and/or lateral forces [7] arise. Based on the above, there have been proposals to rotate or transport a large object with "the force from nothing": Casimir torques and the DLP condition, or lateral Casimir forces and the pulsating ratchet mechanism. Moreover, interesting phenomena have been observed between objects of nontrivial geometries or topographies. Notably, when two corrugated objects are interleaved, there are non-monotonic Casimir forces [8]. Near a perforated thin metal plate, there are regions where an elongated metal particle otherwise attracted by the plate would be repelled by the hole [9]. That has enabled Casimir suspension by rings [10] or by fluid-separated enclosing shells [11]. However, three-dimensional quantum trapping (above a plate, at a given position and height) and low-friction quantum propulsion (over a long distance) of nano-objects, using the Casimir force, are still elusive. The key to them lies in tailoring the lateral Casimir force.

In this letter, we show the existence of a lateral Casimir force near a laterally inhomogeneous plate and demonstrate that, when combined with the DLP repulsive Casimir force, the inhomogeneity-induced lateral Casimir force (ILCF) can laterally stabilize or accelerate nano-objects, depending on the specific inhomogeneity profiles. The Casimir effect within an inhomogeneous fluid has been investigated previously, [12] and corrections to both Casimir energy and stress have been obtained. For lateral Casimir forces, translational symmetry breaking is essential, and thus they are usually studied between corrugated surfaces instead of flat plates, or in a sphere-and-corrugated-plate configuration [13]. In those systems, the lateral force always changes periodically and must reverse within a period. Therefore, lateral local alignments of two objects are expected for stationary systems, and long-distance transport of an object is possible only when the other is oscillating [5]. In comparison, ILCF can be significant over tens of microns, long enough to construct a junction for transport of nano-objects in non-oscillating microfluidic systems. Also, there is more freedom in designing inhomogeneity profiles to implement, including uniform, radial, or angular lateral forces, provided that the micro-scale inhomogeneity of the plate could be experimentally achieved in the sense of an effective medium by mixing materials at nano-scales or simply drilling nano-holes of varying filling factors (usually requiring periods to be much smaller than the characteristic wavelength). The proposed inhomogeneity-induced radial Casimir force, when combined with the DLP repulsive force, can laterally trap an object, producing a stable levitation. Our proposed quantum trapping and propulsion, compared with optical counterparts [14], generate no heat, and thus may find unique applications in manipulating heat-sensitive nano-objects. Throughout this letter we consider the general setup (Fig. 1 (a)) where a golden sphere is above an inhomogeneous plate made of Teflon (PTFE), and both of them are immersed in Bromobenzene (BB). The permittivity of the inhomogeneous plate is modelled by a filling factor f,

$$\tilde{\varepsilon}_{plate} = \varepsilon_{PTFE} + f * (\varepsilon_{BB} - \varepsilon_{PTFE}), \quad (1)$$

where f is a function of x and y, ranging from 0 to 1. All permittivities are obtained from Refs. [10, 15]. All magnetic responses are ignored. A room temperature T=300 K is used.

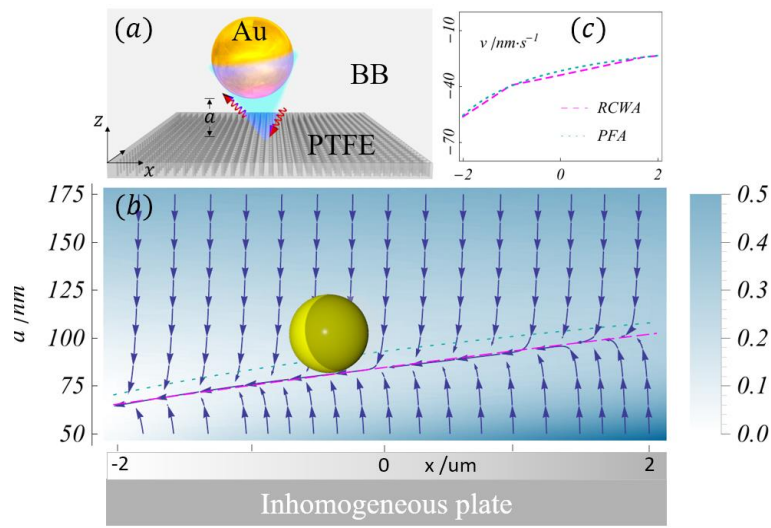

Figure 1 (a) The general setup. Materials of the sphere, plate and fluid are gold, Teflon (PTFE), and Bromobenzene (BB), respectively. Symmetry-breaking of the inhomogeneous plate results in net lateral momentum exchange during reflections, and thus the dominant reflected wave travels in a different direction against the specular one, see red arrows. This results in lateral unbalance of the vacuum radiations (tilted blue light cone), which can be used to drive nano-objects. (b) Schematics of quantum propulsion based on a linearly inhomogeneous (indicated by the grey level) plate. The energy map (normalized by $k_BT$) with force streamlines (also velocity streamlines) is plotted for various lateral positions x (horizontal axis) and separations a (vertical axis) of the sphere. Velocity streamlines show a convergent trajectory (dashed magenta curve), like a slope, that guides the sphere (not to scale) from the right side to the left. Trajectory predicted by proximity approximation is also shown (dotted cyan curve) for comparison. (c) Propulsion velocities along trajectories obtained by rigorous simulations (RCWA, dashed magenta curve) and the proximity force approximation (PFA, dotted cyan curve). Horizontal axis is the position x in microns.

At thermal equilibrium, the free energy of our setup within the scattering formalism reads

$$E = \frac{1}{\beta}\sum_{l=0}^{\infty}{'} Tr \ln\left[\mathbb{I} - \mathcal{R}(\xi_l)\cdot\mathcal{T}(\vec{K}^*,-\vec{a})r(\xi_l)\mathcal{T}(\vec{K},\vec{a})\right], \qquad (2)$$

where $\beta = 1/k_B T$, $k_B$ is Boltzmann's constant, $l$ is the order of Matsubara frequencies $\xi_l = i\frac{2\pi}{\hbar\beta}l$, $\hbar$ is the reduced Planck constant, and the prime above the summation symbol indicates that the $l = 0$ term is weighted by 1/2. Tr is the trace operator that runs over all possible lateral wave numbers k and polarizations $\lambda$. $\mathcal{R}(r)$ is the reflection matrix of the plate (object). $\mathcal{T}(\vec{K},\vec{d}) = \delta_{\lambda,\lambda_0}\delta_{k,k_0}e^{i\vec{K}\cdot\vec{d}}$ is the translation matrix, where $\vec{K} = (k, i\sqrt{k^2 - \varepsilon_3(\xi_l/c)^2})$ and c is the speed of light in vacuum. a is the separation between the object and the plate. In the viewpoint of the object, a virtual shift $\vec{d}$ of itself is equivalent to a virtual shift $-\vec{d}$ of the plate, which can be described by a substitution $\mathcal{R} \rightarrow \mathcal{T}(\vec{K},\vec{d})\mathcal{R}\mathcal{T}(\vec{K}^*,-\vec{d})$. The Casimir force on the object along the d direction, as the negative derivative of the free energy with respect to $d$ when $d \rightarrow 0$, then reads:

$$F = \frac{1}{\beta}\sum_{l=0}^{\infty}{'}\sum_{\lambda,k}\frac{\sum_{\lambda_0,k_0}\left\{[i\vec{K}\tilde{d},\mathcal{R}]_{k,k_0}^{\lambda,\lambda_0}(\mathcal{T}r\mathcal{T})_{k_0,k}^{\lambda_0,\lambda}\right\}}{1 - \sum_{\lambda_0,k_0}\left[\mathcal{R}_{k,k_0}^{\lambda,\lambda_0}(\mathcal{T}r\mathcal{T})_{k_0,k}^{\lambda_0,\lambda}\right]}, \qquad (3)$$

where $[i\vec{K}\tilde{d},\mathcal{R}]_{k,k_0}^{\lambda,\lambda_0} \equiv (i\vec{K}\tilde{d})\cdot\mathcal{R}_{k,k_0}^{\lambda,\lambda_0} - \mathcal{R}_{k,k_0}^{\lambda,\lambda_0}\cdot(i\vec{K}_0^*\tilde{d})$ represents the momentum exchange along the $\tilde{d}$ direction during the reflection process, $\mathcal{T}r\mathcal{T}$ abbreviates $\mathcal{T}(\vec{K}^*,-\vec{a})r\mathcal{T}(\vec{K},\vec{a})$, and the summation over k (k0) indicates integration if k (k0) is continuous. For lateral forces, $[i\vec{K}\tilde{d},\mathcal{R}]$ reduces to $[ik,\mathcal{R}]$. Note that F=$\sum_k$=1/2$\sum_k$+1/2$\sum_{-k}$ and the first term is related to the second one by lateral space inversion, under which $\mathcal{T}r\mathcal{T}$ is symmetric (for symmetric objects) while $[ik,\mathcal{R}]$ is mapped into its complex conjugate $[-ik,\mathcal{R}^*]$. The lateral force must vanish if $\mathcal{R}^* = \mathcal{R}$, i.e., the plate is laterally symmetric. Otherwise, nonzero imaginary part Im($\mathcal{R}$) entailed by symmetry breaking of the plate would cause net lateral momentum exchange and give rise to the lateral Casimir force, as schematically shown in Fig. 1(a).

Numeric simulations based on the rigorous coupled-wave analysis (RCWA) [16] and the supercell technique are carried out to examine the magnitude of ILCF and investigate its applications. Figure 1(b) demonstrates the ILCF-based quantum propulsion of a nano-object. The plate is 4um large, and the filling factor is chosen as $f = \frac{11}{72}(3 - x)$. Due to the spherical symmetry, we are safe to freeze rotational degrees of freedom of the sphere and only consider the translational motion. For different positions and separations of the sphere, the corresponding total energy (free energy plus gravitational potential) is calculated and normalized by $k_B T$. The energy map based on 48 runs of different configurations (a and x), with force streamlines (the lateral components F$_l$ are 3-times exaggerated for better presentation), is shown as the colour map, where the energy minimum has been shifted to zero. Lowest-energy separations (LES, where vertical components of the force streamlines are zero) can be found around 65~105 nm for different lateral positions, revealing quantum levitations of the sphere. The possible friction on the suspended sphere is merely the Stokes drag force $F = \gamma v$, where $\gamma = $

$6\pi\eta r$, η=1.124e-3 Kg/m/s is the viscosity of BB, r is the radius of the sphere, and v is the equilibrium velocity. Therefore, the force streamlines coincide with the equilibrium velocity streamlines, and the latter clearly show a convergent trajectory (dashed magenta curve in Fig. 1(b)) for the sphere; whatever the position and separation are, a sphere inputted at the right side of the plate will move to and later move along the trajectory until it reaches the left side of the plate. Note that the slope of the trajectory is quite small (40nm/4um~1%): it is convenient to treat the trajectory as if it coincided with the LES curve. Further, along the trajectory, the net force gives an averaged propulsion velocity of -34nm/s (dashed magenta curve in Fig. 1(c)), at which it would take 118s for the sphere to cross the inhomogeneous plate. The proximity approximation [17] qualitatively explains the above results. The proximity free energy, as a summation of interacting energy over different face-to-face patches of the real-material surfaces, is calculated by ignoring the dispersion of the materials so that the interacting energy per patch area takes the form of $E_p(z) = -\frac{\hbar c \pi^2}{720 z^3} e$, where the dimensionless factor $e(x) = e_0[1 - f(x)]$ is the ratio of Ep for the considered materials versus that for perfect mirrors. For f in the slow-varying limit, one can arrive at the proximity lateral force

$$F_l = -2\pi R \frac{\hbar c \pi^2}{720 a^3} \times \left(\frac{a f' e0}{2}\right), \qquad (4)$$

and the normal force $F_n = 2\pi R E_p(a)$. Prime means derivative with respect to x. When the DLP condition is imposed, e0<0 and Eq. (4) gives a negative lateral force. The counterbalancing of the normal Casimir force with the gravity of the nano-object implies $a_{LES} \propto \sqrt[3]{(1-f)/R^2}$, predicting a trajectory and its corresponding propulsion velocity shown as the dotted cyan curves in Figs. 1(b) and (c) respectively. Obviously, the proximity approximation predicts the same propulsion phenomena, though differs with RCWA results in magnitudes. Note that the propulsion velocity along the trajectory deduced by Eq. (4) is $v \sim a_{LES}^{-2} \sim R^{4/3}$. It follows that a smaller sphere would experience smaller propulsion velocity, as confirmed by another rigorous simulation based on R=150nm (v150=15nm/s, v270=34nm/s) at x=0um. That might be useful in sorting nano-objects. In addition, $F_l \propto e_0 f'$ means that larger lateral forces are expected for larger inhomogeneities and larger permittivity contrasts. In RCWA simulations, the dimension of the supercell is set to be Lx=Ly=1860nm >>a, and thus the cross interactions between the sphere and its neighbouring cells are significantly scaled down. By comparing with the case of Lx=Ly=1740nm, errors caused by cross interactions are found to be within 1%. 31 orders of spatial harmonics are used to decompose the fields and permittivities. Finite harmonics make the sphere not smooth, but still manage to capture the underlying physics. The first 38 Matsubara frequencies prove sufficient to yield convergent results (errors within 1%), due to the decaying nature of the Matsubara frequency and the low-reflection feature of both objects. The deviations between the PFA results and the RCWA results in Fig. 1 (14% for $a_{LES}$ and 7.1% for v) are mainly ascribed to the violation of a<<R as required by the proximity approximation.

In describing the lateral dynamics of the sphere, Gaussian noise $\xi$ with zero mean and correlations $\langle \xi(t) \cdot \xi(t_0) \rangle = \delta(t - t_0)$ has to be considered in addition to the Stokes drag force, as implied by the fluctuation-dissipation theorem. Since the mass of the sphere $m \ll \gamma$, the non-inertia Langevin equation governs x,

$$\dot{x} = \frac{F_l(x)}{\gamma} + \sqrt{2D}\xi(t), \qquad (5)$$

where dot means derivative with respect to time t, and $D = k_b T/\gamma$ is the well-known Einstein relation. The vertical dynamics is much faster because Fl/Fn < 3%, and thus it is always assumed that the sphere is in vertical equilibrium. The standard deviation of x without Casimir forces is given by $\sigma = \sqrt{2Dt}$. For the case described in Fig. 1, $\sigma(11) = 4um$, that is, it takes 11s for the sphere to cross the plate in a diffusion way, much shorter than the time needed by propulsion and thus the Brownian motion dominates over the propulsion effect. The Casimir force will come to play a leading role for larger spheres. Note that when R is increased, $a_{LES}$ will get smaller, and the proximity approximation would become more precise, though RCWA almost fails due to enormous resource consumptions. By evaluating Eq. (4) along corresponding LES curves and solving Eq. (5) with initial value x=0, Fig. 2 demonstrates the quantum propulsion in the Brownian motion background, for R=2um and f=11/144 (2-x). $a_{LES}$ is found within 20~28 nm. Statistically, the sphere travels 7.0 um in half a minute under the influence of directional lateral Casimir forces, while the diffusion gives $\sigma = 2.4$um. The standard deviation (1σ range) obtained by 300 runs of simulations is marked by shadowed areas around corresponding typical trajectories of motion. Trajectory splitting can be clearly observed after 19s, which exactly demonstrates the quantum propulsion effect.

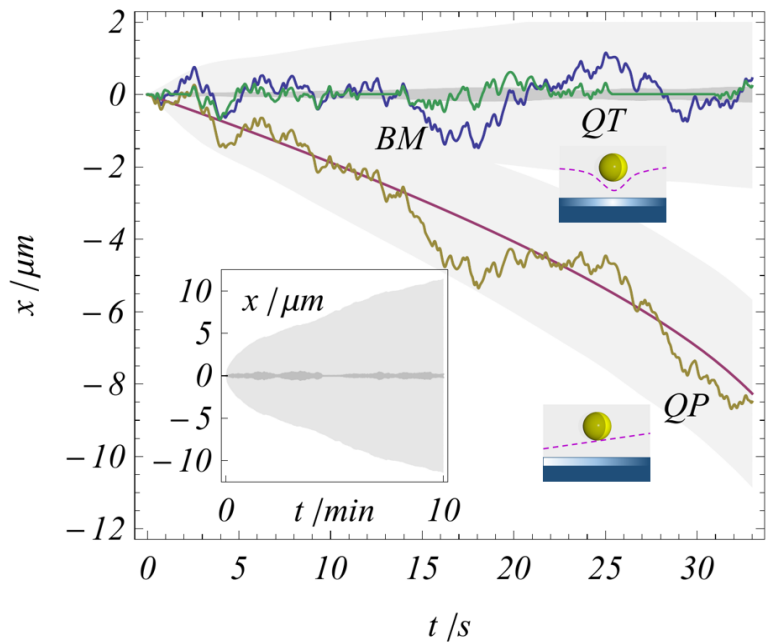

Figure 2 Typical trajectories for quantum propulsion (QP, golden curve) and quantum trapping (QT, green curve) in the Brownian motion (BM, blue curve) background, with corresponding 1σ ranges shadowed. A long-duration test of the standard deviation for BM and QT is given in the lower left inset. Simulating parameters are R=2um, f=11/144 (2-x) (schematically shown in the lower right inset) for QP and f=11/96 (8-|x|) (upper right inset) for QT. The magenta curve represents the propulsion trajectory without Brownian motion.

While a linear (along x direction) inhomogeneity profile has shown its ability to transport nano-objects, one could expect to design a complex inhomogeneity profile in the x-y plane by sophisticated fabrications to guide nano-objects along any desired path or to design a monotonic inhomogeneity profile along the radial direction to trap nano-objects at a fixed height and position. The green curve and heavier shadows in Fig. 2 show features of a sphere (R=2um) trapped by a profile f=11/96 (8-|x|) (a one-dimensional version of radial inhomogeneity is considered for simplicity). $a_{LES}(x = 0)$ is found to be 21nm. A 10min test of the standard deviations of the Brownian motion and quantum trapping is shown in the lower left inset. While $\sigma_{BM} \propto \sqrt{t}$ is observed as expected, $\sigma_{QT}$ almost reaches a plateau at 10s and is of much smaller magnitude (0.5um) than $\sigma_{BM}$. According to the proximity approximation, the ratio of the Casimir energy to the normal Casimir force is a/2. Therefore, one can deduce that the depth of potential well of the quantum trapping $\Delta = \widetilde{m}g * (\frac{\delta a_{LES}}{2} + \delta a_{LES}) \propto$

$e_0^{1/3} R^{7/3}$, $\widetilde{m}$ is the effective mass of the sphere in the fluid. It means larger permittivity contrast is desirable and larger spheres will be trapped more firmly. For the present case, $\Delta = 19.8\, k_b T$. One can further design a slow-varying radial inhomogeneity profile to guide nano-objects far away to the trapping centre, or design a sharp profile to enhance the trapping stiffness.

To show that the smoothly inhomogeneous plate to guide nano-objects can be replaced with a plate full of gratings (a one-dimensional version of nano-holes), two limiting cases are discussed. When $a_{LES} \ll R$ and the proximity approximation applies, $F_l = \int_{-R}^{R} \int_{-\pi/2}^{\pi/2} \sqrt{R^2 - \lambda^2}\, Ep' \cdot d\theta d\lambda$. Note that f is now a square wave function that changes rapidly. Its derivative gives delta functions at the edges of each grating, and thus the lateral force is a sum of contributions near the edges. That makes the force strongly oscillate with magnitude and frequency depending on the period p of gratings, as shown in Fig. 3. The previous propulsion case in Fig. 2 is considered, and p=211, 311, and 411 nm are investigated. For larger p, it can be seen that the oscillating frequency is lower, and the oscillating magnitude is larger. All Fl=0 points on the descending side of the peaks form local equilibrium positions. However, Brownian motion causes the sphere to transit among different local equilibrium positions. A statistical analysis on the probability distribution of x (100nm binned) is performed based on 300 runs of simulations. Results show that strong local equilibria in the p=411nm case almost yield pure quantum trapping effects. The sphere is trapped around a few periods of gratings within half a minute (top left inset). Periods larger than 411nm are also simulated, and they produce better confinement of the sphere (not shown). For p smaller than 211nm, the oscillating magnitude is so small that the lateral Casimir force becomes directional. No local equilibrium guarantees a propulsion effect (bottom inset) almost as good as that based on a smooth inhomogeneity profile. Periods in between (such as 311 nm, top right inset) produce both quantum trapping and quantum propulsion, and thus generally show slowed-down propulsion effects.

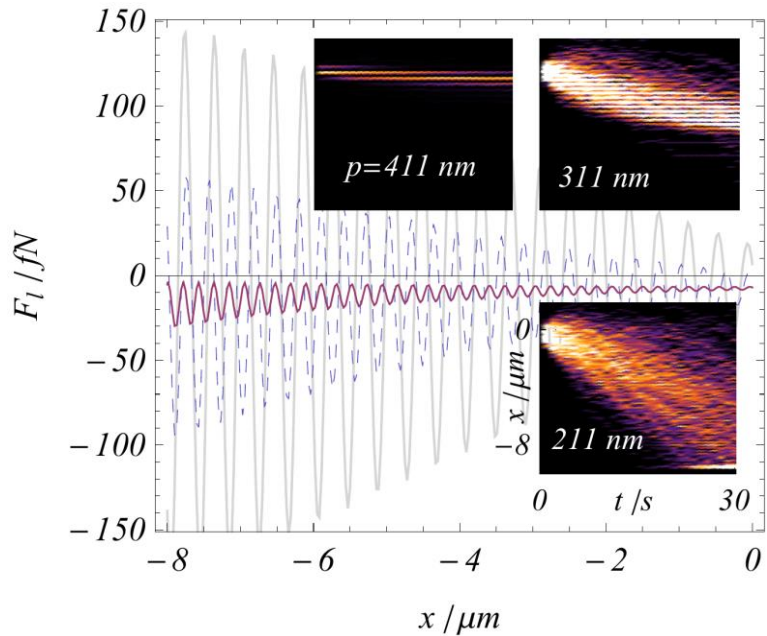

Figure 3 Lateral Casimir forces on a sphere (R=2 um) above plates full of periodic gratings of varying filling factors f=11/144 (2-x). Periods are p=411 nm (grey curve), p=311 nm (blue dashed curve), and p=211 nm (red curve), respectively. Corresponding evolutions of probability distribution of x are given in three insets.

When R is small and $a_{LES} \sim \xi_l/c$, large Kz is overdamped, and thus the Casimir force is determined mainly by small-k contributions (which barely sense local details) and small-$\xi$ contributions (where the effective medium theory applies). RCWA simulations show that, for the case described in Fig. 1 (assuming the sphere be held above the LES curve), p=270nm is sufficient to produce a directional lateral Casimir force when the separation a>343nm. That might be useful in dilute systems where the Brownian motion is negligible.

In summary, quantum propulsion and trapping of nano-objects based on the inhomogeneity-induced lateral Casimir force have been first proposed in the presence of the Brownian motion, and they can be implemented by gratings/nano-holes of varying filling factors. Smooth inhomogeneities can be replaced with varying gratings not only in the effective medium region (a~R), but also in the proximity approximation region (a<<R). The trapping depth is on the order of 10 $k_B$T, and the propulsion distance is over a number of microns. Other material configurations satisfying the DLP condition are also possible to observe those effects. Finite thickness of the inhomogeneous layer weakens those effects, but usually 100nm thick is enough due to the rapid decaying nature of Matsubara frequencies.